\documentclass[12pt,a4paper]{article}\usepackage{a4wide}

\usepackage{epsfig}
\usepackage{amssymb,latexsym,amsmath}
\usepackage{amsfonts}
\usepackage{amsbsy}

\numberwithin{equation}{section}

\DeclareMathOperator{\tr}{tr}
\DeclareMathOperator{\rk}{rank}

\newcommand{\ui}{\textrm{i}}
\newcommand{\ue}{\textrm{e}}
\newcommand{\UI}{\textrm{I}}

\newcommand{\ud}{\mathrm{d}}

\newcommand{\rz}{{\mathbb R}}

\newcommand{\kz}{{\mathbb C}}

\newcommand{\A}{{\mathbb A}}
\newcommand{\B}{{\mathbb B}}
\newcommand{\D}{{\mathbb D}}
\newcommand{\SM}{{\mathbb S}}
\newcommand{\T}{{\mathbb T}}

\begin{document}

\thispagestyle{empty}

\noindent
ULM-TP/02-9\\
October 2002\\

\vspace*{1cm}

\begin{center}

{\LARGE\bf   Spectral Statistics for the Dirac Operator \\
\vspace*{2mm} on Graphs} \\ 
\vspace*{3cm}
{\large Jens Bolte}%
\footnote{E-mail address: {\tt jens.bolte@physik.uni-ulm.de}}
{\large and Jonathan Harrison}%
\footnote{E-mail address: {\tt jon.harrison@physik.uni-ulm.de}}

\vspace*{1cm}

Abteilung Theoretische Physik\\
Universit\"at Ulm, Albert-Einstein-Allee 11\\
D-89069 Ulm, Germany 
\end{center}

\vfill

\begin{abstract}
We determine conditions for the quantisation of graphs using the Dirac 
operator for both two and four component spinors.
According to the Bohigas-Giannoni-Schmit conjecture
for such systems with time-reversal symmetry the energy level statistics are 
expected, in the semiclassical limit, to correspond to those of random 
matrices from the Gaussian symplectic ensemble. 
This is confirmed by numerical investigation.  The scattering matrix 
used to formulate the quantisation condition is found 
to be independent of the type of spinor.  
We derive an exact trace formula for the 
spectrum and use this to investigate the form factor in the diagonal 
approximation.  
\end{abstract}

\newpage

\section{Introduction}
Quantum graphs have proved an important model in the semiclassical 
study of systems whose classical analogues are chaotic
\cite{paper:kottossmilansky2,paper:kottossmilansky}.
Much work in this field concerns the conjecture 
of Bohigas, Giannoni and Schmit 
\cite{paper:bohigasgiannonischmit} which connects statistics of the 
energy level spectrum to those of random matrices, the ensemble of random 
matrices depending on the symmetries of the system.  For systems with no 
time-reversal symmetry statistics of the Gaussian unitary ensemble (GUE)
are expected whereas those of the Gaussian orthogonal (GOE) or symplectic
(GSE) ensemble should apply for systems with time-reversal symmetry, 
depending on whether the spin is integer or half integer, respectively. 
Most investigations of this conjecture have concentrated on systems 
with spin zero and there are few examples of systems whose energy level 
statistics follow the GSE prediction
\cite{paper:scharfdietzkushaakeberry,paper:cauriergrammaticos,paper:scharf,paper:thahablumelsimlansky,paper:kepplermarklofmezzadri}.  
In line with this the usual graph quantisation applies the Schr\"odinger 
operator to a metric graph, finding energy level statistics of the GOE or
GUE \cite{paper:kottossmilansky2,paper:kottossmilansky}.

It is our aim to develop a quantisation of graphs including half integer
spin and we expect to observe spectral statistics following the GSE when
time-reversal symmetry is present.  We choose a Dirac operator that we
realise as a self-adjoint operator on an appropriate Hilbert space.    
The Dirac operator on a graph was considered previously by Bulla and Trenkler 
\cite{paper:bullatrenkler} as an alternative model of a simple scattering 
system, however, without addressing the problem of time-reversal invariance.
Instead we take closed graphs and find boundary conditions that ensure
a self-adjoint realisation of the Dirac operator such that time-reversal 
symmetry is preserved.  By looking at systems of this type we can compare the
spectrum with the results for the Schr\"odinger operator on graphs and general 
semiclassical results for systems with spin.  One well known advantage of 
the quantum graph is that the trace formula is exact rather than 
semiclassical.  This provides the opportunity to distinguish features
present only in the semiclassical limit from those inherent in systems with 
spin.

One unusual quality of the Dirac operator in one dimension is the possibility
of two component spinors rather than the usual four component spinors required
in three dimensions.  Physically this appears odd.  If the Dirac operator on 
the graph describes an idealisation of a physical system of wires in the limit 
that the width of the wire tends to zero we are lead to four component spinors
of the type required in three dimensions.  If however we set up the 
mathematical problem in which the graph is a topological entity then it 
is natural to choose two component spinors.  In the context of quantising 
graphs the apparent contradiction will in fact disappear.  The 
spectrum is independent of the choice of spinors so the physics 
of the system cannot distinguish the language used to describe it.

Our paper is organised as follows: In section \ref{s:graph} we introduce the 
necessary terminology of graphs.  Section \ref{s:dirac} sets out the two 
approaches to the Dirac operator in one dimension, restricting from the Dirac 
equation in three dimensions which leads to four component spinors or 
considering an irreducible representation of the Dirac algebra which requires 
only spinors with two components.  In section \ref{s:2comp} we find 
self-adjoint and time-reversal symmetric realisations of the Dirac operator 
for two component spinors and in section \ref{s:4comp} we do the same in the 
four component case.  Numerical examples confirm that even for simple graphs 
the 
energy level statistics follow the GSE prediction.  We find that the 
scattering matrix in both formalisms has the same properties, the spectrum is 
therefore independent of the approach taken.  In section \ref{s:stats} we 
first derive the exact trace formula for the density of states 
and then use this to calculate the 
form factor.  Making the diagonal approximation we see that the form
factor agrees with 
the GSE result for low $\tau$.  Finally we suggest how the recipe for 
quantising a graph with the Dirac operator is consistent with the model of 
a graph as a three dimensional system of wires, section~\ref{s:exp}.

\section{Graphs}\label{s:graph}
We begin with a few general definitions necessary when considering graphs.
A graph consists of $V$ vertices connected by $B$ bonds.  The valency $v_{i}$ 
of a vertex $i$ is the number of bonds meeting at the vertex.  The topology 
of a graph can be described using its connectivity matrix $C$.  This is 
a $V\times V$ matrix with entries
\begin{equation}\label{eq:1conmatrix}
  C_{ij} := \left\{ \begin{array}{cl}
1 & \textrm{if vertices $i$ and $j$ are connected},\\
0 & \textrm{otherwise},\\
\end{array} \right. 
\end{equation}
where we have assumed that the graph has at most one bond connecting any 
pair of vertices.  We also suppose the graphs under consideration to be 
connected, any vertex can be reached from any other by passing down bonds.  
A graph is directed if the bonds are assigned directions: A bond running from
vertex $i$ to vertex $j$ will be denoted as $b=(ij)$.
 
On a graph one usually considers probabilistic rather than deterministic
classical dynamics.
This can be described by a stochastic matrix $M$ of transition probabilities 
between the bonds.  The matrix $M=(M_{bc})$ defines a Markov chain on the 
graph propagating vectors $\rho=(\rho_1,\dots,\rho_{2B})^T$ of probabilities 
assigned to the bonds, $\rho\mapsto M\rho$.  Note that we allow particles 
to move in either direction on a bond and so there are
$2B$ possible states of the particle on the 
graph.  (This convention for counting bonds will prove necessary later as 
plane-wave solutions for the Dirac operator cannot be made to correspond 
to the directions on a directed graph as in the Schr\"odinger case.)
The non-negative matrix $M$ is
called irreducible if for every pair $(b,c)$ of states there exists a power 
$q$ such that $(M^q)_{bc}>0$.  If the power $q$ can be chosen independent of
$(b,c)$, such that all elements of the matrix $M^q$ are strictly positive, 
$M$ is called primitive.  A Markov chain with an irreducible Markov matrix 
$M$ is ergodic, i.e. any initial distribution $\rho$ converges in average to 
the stable distribution $\rho_0 =\frac{1}{2B}(1,\dots,1)^T$,
\begin{equation}
\lim_{N\to\infty} \frac{1}{N}\sum_{n=1}^N M^n \rho = \rho_0 \ .
\end{equation}
A Markov matrix corresponding to an ergodic chain has a non-degenerate
eigenvalue one.  For primitive stochastic matrices all other eigenvalues
are smaller in magnitude such that the associated Markov process is mixing.
Then an initial distribution converges to the stable distribution,
\begin{equation}
\lim_{n\to\infty} M^n \rho = \rho_0 \ ,
\end{equation}
and, moreover, 
\begin{equation}
\lim_{n\to \infty}  (M^{n})_{bc} = \frac{1}{2B} \ . 
\end{equation}
It is this last condition that will be used in section \ref{ss:5.2} 
to evaluate the form factor in the diagonal approximation.  Details
on non-negative matrices and Markov chains can be found in 
\cite{book:gantmacher}.

If the quantum system on the graph is defined by a matrix $\T=(T_{bc})$ of 
complex transition amplitudes between the bonds (we will see that $\T$ is 
unitary) then 
\begin{equation}
M_{(ij)(ki)}=|T_{(ij)(ki)}|^2 \ .
\end{equation}
$M_{(ij)(ki)}$ is the quantum mechanical probability that a particle 
traveling from $k$ to $i$ scatters at $i$ to travel towards the 
vertex $j$.
The unitarity of $\T$ 
implies that $M$ is doubly stochastic and therefore defines a Markov chain.    

While only topological information is necessary to investigate classical 
dynamics on a graph quantum mechanics requires us to assign lengths, 
$L_{(ij)}$, to bonds, $(ij)$.  This defines a metric graph.  It is natural 
to consider a bond with length as directed, one vertex lying at zero on the 
bond and the other at $L$, consequently our metric graphs are also directed 
graphs.   
In order to avoid degeneracies in the lengths of periodic orbits on the graph 
we assume the lengths assigned to the bonds are incommensurate, not related 
by a rational number.

\section{The Dirac equation in one dimension}\label{s:dirac}
In one spatial dimension the Dirac equation is 
\begin{equation}\label{eq:3dir}
\ui \hbar \frac{\partial}{\partial t} \Psi(x,t) 
= \left( -\ui \hbar c \, \alpha \, \frac{\partial}{\partial x} + 
mc^{2}\, \beta \, \right) \Psi(x,t) \ ,
\end{equation}
where $\alpha$ and $\beta$ satisfy the relations $\alpha^2=\beta^2=\UI$ 
and $\alpha\beta+\beta\alpha=0$ that define the Dirac algebra.  
There are two possible interpretations of this equation:  Either one
views it as a restriction of the Dirac equation in three dimensions,
or one considers it as a problem in one dimension from the outset.  In the 
first case this equation originates from an implementation of the Poincar\'e 
space-time symmetries in relativistic quantum mechanics such that the
notions of spin and of anti-particles can be carried over.  If then,
for instance, one restricts to the $y$ axis the Dirac matrices 
$\alpha=\alpha_y$ and $\beta$ are hermitian $4\times 4$ matrices that form 
a reducible representation of the Dirac algebra in one dimension.  The 
second case is void of the physical interpretations deriving from 
$3+1$-dimensional space-time symmetries.  One merely considers the Dirac
equation (\ref{eq:3dir}) with a faithful 
irreducible representation of the Dirac 
algebra and thus chooses $\alpha$ and $\beta$ to be hermitian $2\times 2$ 
matrices.  For an extensive discussion of the Dirac equation see
\cite{book:thaller}. 

The two cases are naturally connected as a unitary transformation 
of $\alpha_{y}$ and $\beta$ brings them into block diagonal form preserving 
the algebraic relations.  For example the standard (Dirac) representation 
of the matrices in three dimensions, restricted to the $y$ axis, is
\begin{equation}\label{eq:3dirmat}
\alpha = \left( \begin{array}{cccc}
0 & 0 & 0 & -\ui \\
0 & 0 & \ui & 0 \\
0 & -\ui & 0 & 0 \\
\ui & 0 & 0 & 0 \\
\end{array} \right) \ , 
\qquad
\beta = \left( \begin{array}{cccc}
1 & 0 & 0 & 0 \\
0 & 1 & 0 & 0 \\
0 & 0 & -1 & 0 \\
0 & 0 & 0 & -1 \\
\end{array} \right). 
\end{equation}
Now consider the unitary transformation 
\begin{equation}
U = \frac{1}{\sqrt{2}} \left( \begin{array}{cccc}
1 & 1 & 0 & 0 \\
0 & 0 & -1 & 1 \\
-1 & 1 & 0 & 0 \\
0 & 0 & 1 & 1 \\
\end{array} \right).
\end{equation}
Applying $U$ to $\alpha$ and $\beta$ generates a new representation,
\begin{equation}
U\alpha U^{-1} = \left( \begin{array}{cccc}
0 & -\ui & 0 & 0 \\
\ui & 0 & 0 & 0 \\
0 & 0 & 0 & \ui \\
0 & 0 & -\ui & 0 \\
\end{array} \right) 
\ , \qquad
U\beta U^{-1} = \left( \begin{array}{cccc}
1 & 0 & 0 & 0 \\
0 & -1 & 0 & 0 \\
0 & 0 & 1 & 0 \\
0 & 0 & 0 & -1 \\
\end{array} \right) \ , 
\end{equation}
which is obviously reducible. Either of the $2\times 2$ blocks 
represents the Dirac algebra in one dimension irreducibly. 
It is, however, important to note that passing from the four dimensional  
to the two dimensional representation the usual interpretations of spin, 
particles and anti-particles, and time-reversal are lost. 

For both two- and four-dimensional representations of the Dirac
algebra the Dirac operator reads
\begin{equation}\label{eq:3dop}
\mathcal{D}
:= -\ui \hbar c \, \alpha \, \frac{\ud}{\ud x} + 
mc^{2}\, \beta \ .
\end{equation}
On the real line $\mathcal{D}$ is defined as an essentially self-adjoint 
operator on the Hilbert space $L^2(\rz)\otimes\kz^n$ with 
domain $C_0^\infty(\rz)\otimes\kz^n$, where $n=2$ or $n=4$.

In order to obtain self-adjoint realisations of the Dirac operator on
a compact interval $I=[0,L]$ appropriate boundary conditions have to be 
specified. These can be classified by extensions of the closed symmetric 
operator $\mathcal{D}^0$ given as $\mathcal{D}$ defined on the domain 
$W_{2,1}^0(I)\otimes\kz^n \subset L^2(I)\otimes\kz^n$.  Here $W_{2,1}^0(I)$ 
denotes the Sobolev space of functions $\varphi$ on $I$ that are, along 
with their (generalised) first derivatives, in $L^2(I)$ and satisfy 
$\varphi(0)=0=\varphi(L)$.  The closed symmetric extensions of $\mathcal{D}^0$
arise from restrictions of $\mathcal{D}$ defined on the domain 
$W_{2,1}(I)\otimes\kz^n$ of $n$-component Sobolev spinors with no specified
boundary conditions to subspaces on which the skew-Hermitian quadratic form 
\begin{equation}\label{eq:3form}
\Omega(\phi,\psi):=\langle \mathcal{D} \phi, \psi \rangle - \langle \phi, 
\mathcal{D} \psi \rangle
\end{equation}
vanishes.  The maximal isotropic subspaces of $W_{2,1}(I)\otimes\kz^n$
with respect to $\Omega$, i.e. the maximal subspaces on which $\Omega$ 
vanishes, then yield domains on which $\mathcal{D}$ is self-adjoint. 
This is the approach adopted by Kostrykin and Schrader for the 
Schr\"odinger operator \cite{paper:kostrykinschrader} on graphs.  
Self-adjoint realisations of the Dirac operator on an interval
are also classified using a different technique by Alonso and De Vincenzo 
\cite{paper:alonsovincenzo}.

\section{The Dirac operator for two component spinors}\label{s:2comp}
From the two approaches to the problem we begin with the minimum necessary
and consider two component spinors first.  According to the above we take
\begin{equation}\label{eq:3alphabeta}
\alpha = \left( \begin{array}{cc}
0 & -\ui \\
\ui & 0 \\
\end{array} \right) \ ,
\qquad
\beta = \left( \begin{array}{cc}
1 & 0 \\
0 & -1 \\
\end{array} \right) .
\end{equation}
(The choice of $\alpha$ and $\beta$ is designed to simplify the calculations.)

Eigenspinors $\psi=\binom{\psi_1}{\psi_2}$ of $\mathcal{D}$ are solutions 
of the time independent Dirac equation 
\begin{equation}\label{eq:3tdir}
\mathcal{D}  \psi(x)=E \psi(x)\ .
\end{equation}
For positive energies $E$ they are hence plane waves of the form 
\begin{equation}\label{eq:3planewave}
\psi_k(x) = \mu(k) \left( \begin{array}{c} 
1 \\
\ui \gamma(k)\\
\end{array}\right) \ue^{\ui k x} +
\hat{\mu}(k) \left( \begin{array}{c} 
1 \\
-\ui \gamma(k) \\
\end{array}\right) \ue^{-\ui k x}
\end{equation}
with $k > 0$ and
\begin{equation}\label{eq:3dispersion}
\gamma(k) := \frac{E-mc^2}{\hbar c k} \ ,\qquad 
E= \sqrt{ (\hbar c k)^{2} +m^{2}c^{4} } \ .
\end{equation}
The choice of the coefficients $\mu(k)$ and $\hat\mu(k)$ depends on the
boundary conditions imposed by the self-adjoint realisation of $\mathcal{D}$.
We see immediately that the plane-wave spinor (\ref{eq:3planewave}) is not 
invariant changing $x$ to $-x$.  As the Dirac operator is first order the 
direction assigned to bonds on the graph becomes significant.

\subsection{Self-adjoint realisations on graphs}
Self-adjoint realisations of the Dirac operator on graphs were considered by 
Bulla and Trenkler \cite{paper:bullatrenkler}.   However, in order to develop 
a simple form related to the vertex transition matrix, we continue with
the approach of Kostrykin and Schrader \cite{paper:kostrykinschrader} for the 
Schr\"odinger operator.

The basic idea is to view a graph as a collection of intervals that are
glued together according to the connectivity matrix $C$. The Hilbert
space on which the Dirac operator acts is therefore the direct sum of the
Hilbert spaces for each bond,
\begin{equation}
{\mathcal H} := \bigoplus_{b=1}^B L^2 \bigl( [0,L_b] \bigr)
\otimes\kz^2 \ ,
\end{equation}
where $b$ runs over all bonds. The spinors 
$\psi\in{\mathcal H}$ therefore consist of the $B$ components
$(\psi^{1},\dots ,\psi^{B})$ attached to the bonds, each of which is a 
two-spinor. The scalar product of spinors on the graph is then
\begin{equation}
\langle \phi , \psi \rangle = 
\sum_{b=1}^{B} \langle \phi^{b}, \psi^{b} \rangle_b \ , 
\end{equation}
where $\langle \phi^{b}, \psi^{b} \rangle_b$ is the scalar product in 
$L^2 \bigl( [0,L_b] \bigr) \otimes\kz^2$.
Self-adjoint realisations of $\mathcal{D}$ are constructed in analogy to 
the case of a single interval by suitably extending the closed symmetric 
operator $\mathcal{D}^0$ defined on the domain 
\begin{equation}
\bigoplus_{b=1}^B W^0_{2,1} \bigl( [0,L_b] \bigr) \otimes\kz^2 \ . 
\end{equation}
The domains of its closed symmetric extensions are the isotropic subspaces
of 
\begin{equation}\label{eq:3Sobolevspace}
\bigoplus_{b=1}^B W_{2,1} \bigl( [0,L_b] \bigr) \otimes\kz^2  
\end{equation}
with respect to the skew-Hermitian form $\Omega$ as given in
(\ref{eq:3form}). Integrating by parts yields
\begin{equation} 
\Omega(\phi,\psi)= 
\hbar c \sum_{b=1}^{B} \Bigl( \psi^{b}_{1}(0)\overline{\phi}^{b}_{2}(0)
-\psi^{b}_{1}(L_b)\overline{\phi}_{2}^{b}(L_b)
-\psi^{b}_{2}(0)\overline{\phi}_{1}^{b}(0)
+\psi^{b}_{2}(L_b)\overline{\phi}_{1}^{b}(L_b) \Bigl) \ ,
\end{equation}
showing that $\Omega$ only depends on the boundary values of the
spinor components.  Self-adjoint extensions again correspond to maximally 
isotropic subspaces and thus can be characterised by boundary conditions.

Following the construction of Kostrykin and Schrader 
\cite{paper:kostrykinschrader} for Schr\"odinger operators on graphs, we
introduce a map from the infinite-dimensional Hilbert space ${\mathcal H}$
to the $4B$-dimensional space of boundary values, $\psi \mapsto 
\boldsymbol{\psi} = \left(
\begin{smallmatrix}\boldsymbol{\psi}_{1} \\ \boldsymbol{\psi}_{2}
\end{smallmatrix} \right)$, with
{\setlength\arraycolsep{2pt}
\begin{equation}
\begin{split}
\boldsymbol{\psi}_{1}&:=\bigl( \psi_{1}^{1}(0),\dots,\psi_{1}^{B}(0), 
\psi_{1}^{1}(L_{1}), \dots ,\psi_{1}^{B}(L_{B}) \bigr)^{T} \ , \\
\boldsymbol{\psi}_{2}&:=\bigl( -\psi_{2}^{1}(0),\dots,-\psi_{2}^{B}(0), 
\psi_{2}^{1}(L_{1}), \dots ,\psi_{2}^{B}(L_{B}) \bigr)^{T} \ .
\end{split}
\end{equation}
}By including $\psi_{2}(0)$ with a minus sign we can write $\Omega$ as 
\begin{equation}\label{eq:3omegascalar}
  \Omega(\phi,\psi)= ( \begin{array}{cc}
  \boldsymbol{\phi}_{1}^{\dagger}& \boldsymbol{\phi}_{2}^{\dagger} \\
\end{array} ) \left( \begin{array}{cc}
0 & \UI_{2B} \\
-\UI_{2B} & 0 \\
\end{array} \right) \left( \begin{array}{c}
\boldsymbol{\psi}_{1}\\
\boldsymbol{\psi}_{2}\\
\end{array} \right) \ .
\end{equation}
The maximal isotropic  subspaces in (\ref{eq:3Sobolevspace}) with respect
to $\Omega$ are now equivalent to maximal subspaces of vectors 
$\boldsymbol{\psi} \in \mathbb{C}^{4B}$ on which the complex 
symplectic form (\ref{eq:3omegascalar}) vanishes.  Defining a linear 
subspace of $\mathbb{C}^{4B}$ as the vectors $\boldsymbol{\psi}$ satisfying
\begin{equation}\label{eq:3boundarycond}
\A \boldsymbol{\psi}_{1} + \B \boldsymbol{\psi}_{2} = 0 \ ,  
\end{equation}
with complex $2B\times 2B$ matrices $\A$ and $\B$, one first observes that 
in order for this subspace to have the desired dimension $2B$ the 
$2B\times 4B$ matrix $(\A,\B)$ must have maximal rank. Kostrykin and Schrader
show that such a subspace is maximally isotropic if and only if $(\A,\B)$ 
has maximal rank and $\A\B^{\dagger}$ is hermitian.  

Equation (\ref{eq:3boundarycond}) defines boundary conditions for the Dirac 
operator on a collection of $B$ bonds.  These boundary conditions yield 
a self-adjoint realisation of ${\mathcal D}$ when 
\begin{equation}\label{eq:sae}
\rk(\A,\B)=2B \quad \text{and} \quad \A\B^{\dagger}=\B\A^{\dagger} \ .
\end{equation}

So far the connectivity of the graph has played no role. In order to 
implement this we consider only boundary conditions, specified through
the matrices $\A$ and $\B$, that connect bonds according to the connectivity
matrix $C$. The matrices $\A$ and $\B$ then have a block structure, each block 
defining boundary conditions at a single vertex.

\begin{figure}[htb]
\begin{center}
\includegraphics[width=4cm]{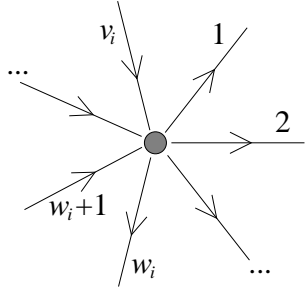}
\end{center}
\caption{An isolated vertex on which we wish to define the transition matrix.}
\label{fig:3star}
\end{figure}

At a given vertex 
$i$ we label the $v_{i}$ bonds such that the first $w_{i}$ bonds are 
outgoing and the remainder incoming, see figure \ref{fig:3star}.  The vector
$\boldsymbol{\psi}^{(i)} = \left(
\begin{smallmatrix}\boldsymbol{\psi}_{1}^{(i)}  \\ \boldsymbol{\psi}_{2}^{(i)} 
\end{smallmatrix} \right) \in \kz^{2v_i}$ of boundary values at $i$
then consists of the components
{\setlength\arraycolsep{2pt}
\begin{equation}
\begin{split}
\boldsymbol{\psi}_{1}^{(i)}
&=\bigl( \psi_{1}^{1}(0),\dots,\psi_{1}^{w_i}(0), 
\psi_{1}^{w_{i}+1}(L_{w_{i}+1}), \dots ,
\psi_{1}^{v_i}(L_{v_i}) \bigr)^{T} \ ,\\
\boldsymbol{\psi}_{2}^{(i)}
&=\bigl( -\psi_{2}^{1}(0),\dots,-\psi_{2}^{w_i}(0), 
\psi_{2}^{w_{i}+1}(L_{w_{i}+1}), \dots ,
\psi_{2}^{v_i }(L_{v_i }) \bigr)^{T} \ .
\end{split}
\end{equation}
}If we denote the $v_i\times v_i$ blocks of the matrices $\A$ and $\B$ that
define the boundary conditions at $i$ by $\A^{(i)}$ and $\B^{(i)}$, we have 
the condition
\begin{equation}\label{eq:3boundarycond2}
\A^{(i)} \boldsymbol{\psi}_{1}^{(i)}+\B^{(i)} \boldsymbol{\psi}_{2}^{(i)} = 0 
\quad \textrm{ with } \quad  \A^{(i)}\B^{(i)\dagger}=\B^{(i)}\A^{(i)\dagger} 
\end{equation}
which guarantees current conservation at the vertex $i$.

Eigenspinors $\psi$ of the Dirac operator ${\mathcal D}$ with given
boundary conditions are composed of plane waves, equation 
(\ref{eq:3planewave}), on each of the bonds.  Assigning incoming and 
outgoing plane-wave solutions to each of the bonds we can write down 
vectors of coefficients $\boldsymbol{\overrightarrow{\mu}}$, 
$\boldsymbol{\overleftarrow{\mu}}$ for the outgoing and incoming waves at 
the vertex, respectively:
\begin{equation}
\boldsymbol{\overrightarrow{\mu}}=\left( \begin{array}{c}
\mu^{1}\\
\vdots \\
\mu^{w_{i}}\\
\hat{\mu}^{w_{i}+1}\,\ue^{-\ui kL_{w_{i}+1}}\\
\vdots \\
\hat{\mu}^{v_{i}}\,\ue^{-\ui kL_{v_{i}}}
\end{array} \right) \ , \qquad 
\boldsymbol{\overleftarrow{\mu}}=\left( \begin{array}{c}
\hat{\mu}^{1}\\
\vdots \\
\hat{\mu}^{w_{i}}\\
\mu^{w_{i}+1}\,\ue^{\ui kL_{w_{i}+1}}\\
\vdots \\
\mu^{v_{i}}\,\ue^{\ui kL_{v_{i}}}\\
\end{array} \right) \ .
\end{equation}
Here we omit the index $i$ labeling the vertex.
Writing $\boldsymbol{\psi}_{1}^{(i)}$ and $\boldsymbol{\psi}_{2}^{(i)}$ 
in terms of $\boldsymbol{\overrightarrow{\mu}}$ and 
$\boldsymbol{\overleftarrow{\mu}}$ and substituting 
into the definition of the boundary conditions we obtain
\begin{equation}
\boldsymbol{\overrightarrow{\mu}}=-(\A^{(i)} + \ui \gamma (k) \B^{(i)})^{-1}
(\A^{(i)} - \ui \gamma (k) \B^{(i)})\, 
\boldsymbol{\overleftarrow{\mu}} \ .
\end{equation}
Notice that under the conditions given the matrices 
$(\A^{(i)} \pm \ui \gamma (k) \B^{(i)})$ are always invertible 
\cite{paper:kostrykinschrader}.
This defines the vertex transition matrix $\T^{(i)}$ which connects the 
coefficients of incoming and outgoing plane waves at the vertex $i$,
\begin{equation}\label{eq:3transition}
 \T^{(i)} :=-(\A^{(i)} + \ui \gamma (k) \B^{(i)})^{-1}
(\A^{(i)} - \ui \gamma (k) \B^{(i)}) \ .
\end{equation}
The condition that $\A^{(i)}\B^{(i)\dagger}$ is hermitian implies that 
the matrix $\T^{(i)}$ is unitary,
\begin{equation}
  (\T^{(i)})^{-1}=(\T^{(i)})^{\dag} \ .
\end{equation}
Self-adjoint realisations of the Dirac operator on the graph are defined by 
matrices $\A,\B$ with $\A\B^{\dagger}$ hermitian.  These prescribe the 
boundary conditions (\ref{eq:3boundarycond}) and determine unitary 
vertex transition matrices (\ref{eq:3transition}) for plane waves on 
the graph.

\subsection{Time-reversal symmetry}
Following the conjecture of Bohigas, Giannoni and Schmit,
we expect to find energy level statistics corresponding to those of the 
Gaussian symplectic ensemble (GSE) for quantum systems with half-integer 
spin $s$ and time-reversal symmetry.  This is due to the fact that the 
time-reversal operator is anti-unitary and squares to $(-1)^{2s}$, imprinting 
a symplectic symmetry on the Hamiltonian.  In the case of a two component
Dirac equation in one dimension the usual interpretations
of spin and time-reversal inherited from three dimensions are lost.  
Nevertheless, one can introduce an anti-unitary operator squaring to $-1$ 
that is formally identical to the time-reversal operator for two component 
(Pauli-) spinors in three dimensions,  
\begin{equation}\label{eq:3timeop}
\mathcal{T}:=
\left( \begin{array}{cc}
0 & 1 \\
-1 & 0 \\
\end{array} \right)
\mathcal{K} \ ,
\end{equation}
where $\mathcal{K}$ is complex conjugation. This can be interpreted as 
a physical time-reversal in the sense that it changes the sign of time and 
momentum.  The matrix part of $\mathcal{T}$ is only required in order that
$\mathcal{T}^2=-\UI$ as there is no spin operator for two component spinors.
In the subsequent work we will therefore refer to (\ref{eq:3timeop}) as 
a time-reversal operator.   A full discussion of time-reversal symmetry with 
spin is found in the second chapter of \cite{book:haake}.

We now determine conditions under which the vertex 
transition matrix is time-reversal symmetric.  This will raise questions 
about when general 
time-reversal symmetric boundary conditions 
exist for a Dirac operator on a graph.    For the system to be 
time-reversal symmetric $\mathcal{T}$ must commute with the Hamiltonian 
$\mathcal{D}$.  This forces the mass term in $\mathcal{D}$ to vanish.  
From now on we consider only the case $m=0$ for which 
$\gamma(k)=1$ and time-reversal symmetry is possible with two component 
spinors.

Applying $\mathcal{T}$ to the plane-wave solutions on a bond 
(\ref{eq:3planewave}) we find
\begin{equation}\label{eq:3tplanewave}
\mathcal{T} \psi_k(x) = -\ui \, \overline{\mu} \left( \begin{array}{c} 
1 \\
-\ui \\
\end{array}\right) \ue^{-\ui k x} +
\ui \, \overline{\hat{\mu}} \left( \begin{array}{c} 
1 \\
\ui \\
\end{array}\right) \ue^{\ui k x} \ .
\end{equation}
Time-reversing spinors on all the bonds meeting at the vertex $i$ defines
new vectors of outgoing and incoming waves 
$\boldsymbol{\overrightarrow{\mu}}_{\mathcal{T}}, 
\boldsymbol{\overleftarrow{\mu}}_{\mathcal{T}}$,
\begin{equation}
\boldsymbol{\overrightarrow{\mu}}_{\mathcal{T}} = \ui 
\left( \begin{array}{cc} 
\UI_{w_{i}} &  0 \\
0 & - \UI_{(v_{i}-w_{i})} \\
\end{array} \right) \overline{(\boldsymbol{\overleftarrow{\mu}})} \ ,
\qquad 
\boldsymbol{\overleftarrow{\mu}}_{\mathcal{T}} = \ui 
\left( \begin{array}{cc} 
-\UI_{w_{i}} & 0 \\
0 & \UI_{(v_{i}-w_{i})} \\
\end{array} \right) 
\overline{(\boldsymbol{\overrightarrow{\mu}})} \ .
\end{equation}
For the vertex transition matrix to be time-reversal invariant we require
\begin{equation}\label{eq:3timereversecond}
\boldsymbol{\overrightarrow{\mu}}_{\mathcal{T}} = \T^{(i)} \, 
\boldsymbol{\overleftarrow{\mu}}_{\mathcal{T}} \ .
\end{equation}
Using the definition of the vertex transition matrix, 
$\boldsymbol{\overrightarrow{\mu}}= \T^{(i)} \boldsymbol{\overleftarrow{\mu}}$ 
for $\T^{(i)}$ unitary, equation (\ref{eq:3timereversecond}) implies
\begin{equation}\label{eq:3trcond2}
(\T^{(i)})^{T} = \left( \begin{array}{cc} 
\UI_{w_{i}} & 0 \\
0 & - \UI_{(v_{i}-w_{i})} \\
\end{array} \right)
\T^{(i)}
\left( \begin{array}{cc} 
-\UI_{w_{i}} & 0 \\
0 & \UI_{(v_{i}-w_{i})} \\
\end{array} \right) \ .
\end{equation}
Equation (\ref{eq:3trcond2}) is the condition that all vertex transition 
matrices must satisfy 
in order that the system possesses time-reversal symmetry.

Splitting $\T^{(i)}$ into four blocks equation (\ref{eq:3trcond2}) 
is equivalent to
\begin{equation} \label{eq3trc3}
\T^{(i)}=
\left( \begin{array}{cc} 
\T_{1} & \T_{2} \\
\T_{3} & \T_{4} \\
\end{array} \right) =
\left( \begin{array}{cc} 
-\T^{T}_{1} & \T^{T}_{3} \\
\T^{T}_{2} & -\T^{T}_{4} \\
\end{array} \right) \ .
\end{equation}
The components $T_{bc}$ of the transition matrix $\T^{(i)}$ are thus either 
symmetric or antisymmetric according to the alignment of the bonds at 
the vertex, in the sense that:
\begin{center}
\begin{tabular}{ccccc}
\raisebox{0.5cm}{$T_{bc}=T_{cb}$} & & 
\includegraphics[width=2.5cm]{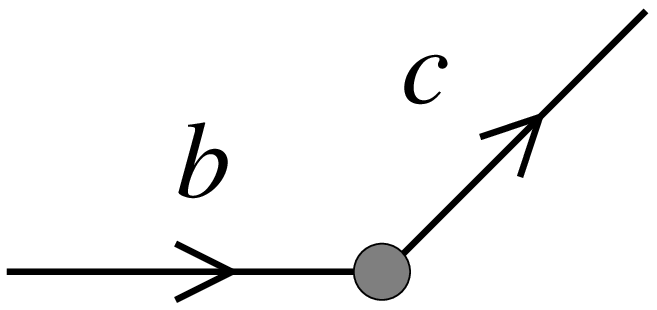} & 
& \\
\raisebox{0.5cm}{$T_{bc}=-T_{cb}$} & & 
\includegraphics[width=2.5cm]{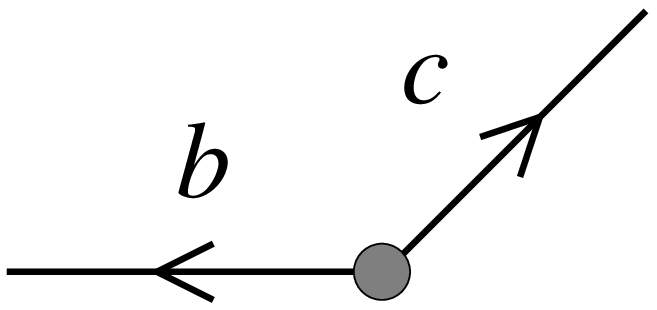} 
& \raisebox{0.5cm}{or} & \includegraphics[width=2.5cm]{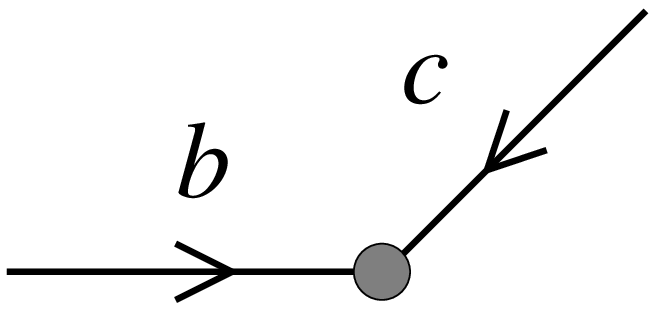} \\
\end{tabular} 
\end{center}
This raises a number of problems.  Back scattering is not possible in this 
scheme as $T_{bb}$ is identically zero (returning to the same bond always
requires antisymmetry).  Consequently it is not possible to quantise graphs 
containing vertices with valency one.  It is also clear that the components 
of $\T^{(i)}$ cannot be invariant under a permutation of the bonds.  But
it is this natural physical property imposed by Kottos and Smilansky 
\cite{paper:kottossmilansky2,paper:kottossmilansky} that allows them to 
derive a simple form of the trace formula 
for the Schr\"odinger operator on a graph.  In fact we may ask if any graphs 
exist for which the vertex transition matrices are both unitary and satisfy 
the time reversal condition (\ref{eq:3trcond2}).  There are no examples of 
$\T^{(i)}$ for vertices with valency one and only a trivial solution for one 
incoming and one outgoing bond, $v_{i}=2$.  With $v_{i}=3$ there are again 
no solutions.  The simplest nontrivial example of a transition matrix is for 
a vertex with valency four.  In this case, with two incoming and two outgoing 
bonds, one possible choice of a time-reversal invariant transition matrix is
\begin{equation}\label{eq:3Teg1}
\T^{(i)}= \frac{1}{\sqrt{2}}
\left( \begin{array}{cccc}
0 & 1 & 1 & 0 \\
-1 & 0 & 0 & 1 \\
1 & 0 & 0 & 1 \\
0 & 1 & -1 & 0 \\
\end{array}
\right) \ .
\end{equation}
Clearly this is both a unitary matrix and has the symmetry and antisymmetry 
properties prescribed by (\ref{eq3trc3}).  Physically this would represent
strange boundary conditions as a spinor reaching the vertex would be 
prevented not only from back scattering but also from scattering down another 
of the bonds.  However it is still reasonable to ask how closely 
the spectrum of
a graph  quantised with such a vertex transition matrix will follow 
the GSE prediction.  In the following subsection we investigate one example 
of a binary graph (i.e. where all vertices have two incoming and two outgoing 
bonds).

\subsection{Bond scattering matrix}
To find the energy levels we must first determine the bond scattering 
matrix $\SM$ for the whole graph from the transition matrices at the vertices.
Let us define $\mu^{(ij)}$ to be the coefficient of the plane wave on the 
bond $(ij)$ traveling in the direction $i\rightarrow j$.  Then $\mu^{(ji)}$ 
is $\hat{\mu}^{(ij)}$ in our previous notation.  The bond scattering matrix 
$\SM=(S_{(ij)(lm)})$ is defined by
\begin{equation}\label{eq:3S}
S_{(ij)(lm)}(k)= \delta_{mi} \ T^{(m)}_{(ij)(lm)} \ \ue^{\ui kL_{(lm)}} \ .
\end{equation}
To scatter from the spinor with coefficient $\mu^{(lm)}$ to that with 
$\mu^{(ij)}$ the bonds $(lm)$ and $(ij)$ must be connected at $m$.  
The transition amplitude is then 
defined by the transition matrix $\T^{(m)}$ at $m$. 
Before the transition the spinor collects a phase
propagating along $(lm)$.  Equation (\ref{eq:3S}) 
is equivalent to the description of 
the scattering matrix as a product $\SM(k)=\T\,\D(k)$ where
$\D(k)$ is a diagonal matrix of phases and $\T$ is a matrix of transition 
amplitudes for the whole graph.  As there are two coefficients for the 
spinors on each bond $\SM$ is a $2B\times 2B$ matrix.  The unitarity of 
the vertex transition matrices and the diagonal matrix of phases $\D(k)$ 
ensures that $\SM(k)$ is also unitary.  Note that the vertex transition 
matrices $\T^{(m)}$ and hence the bond S-matrix still depend on the directions 
assigned to the bonds.

The bond S-matrix $\SM(k)$ acts on the vector of coefficients 
$\boldsymbol{\mu}$, which is composed of the coefficients
$\boldsymbol{\overrightarrow{\mu}}$ and $\boldsymbol{\overleftarrow{\mu}}$
at every vertex and therefore defines the plane wave on the graph.  
Energy eigenfunctions correspond to vectors $\boldsymbol{\mu}$ where
\begin{equation}
\SM(k) \boldsymbol{\mu} = \boldsymbol{\mu} \ .
\end{equation}
The energy eigenvalues then correspond to the values of $k$ for which 
\begin{equation}\label{eq:3quant}
\bigl| \UI_{2B} - \SM(k) \bigr| = 0 \ .
\end{equation}
This quantisation condition for the graph is the same as that for
the Schr\"odinger operator, see 
\cite{paper:kottossmilansky2,paper:kottossmilansky}.

\begin{figure}[htb]
\begin{center}
\includegraphics[width=14cm]{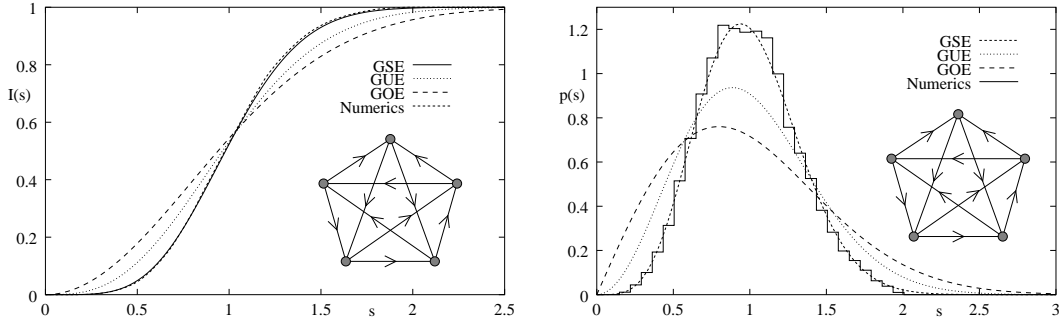}
\end{center}
\caption{The integrated nearest neighbour spacing distribution (left)
and a histogram of the spacing distribution (right) for a fully 
connected Dirac pentagon, with 35753 levels.}
\label{fig:3pent}
\end{figure}

Figure \ref{fig:3pent} shows the nearest neighbour level 
spacing statistics for a fully connected 
pentagon calculated using the vertex transition matrices (\ref{eq:3Teg1}).  
Directions were assigned to the bonds to produce
two incoming and two outgoing bonds at 
each vertex.  The assignment of $\T^{(i)}$ to the vertices is still not 
unique as the elements of $\T^{(i)}$ also vary between bonds with the 
same direction.  The numerics confirm that, even for this small graph with 
unusual boundary conditions, the energy level statistics correspond well
to those of random matrices from the GSE.  We now turn to consider 
how the quantisation of more general graphs can be realised using
two component spinors
and whether there exists a trace formula for the Dirac operator.

\subsection{Paired bonds}
We expect it should be possible to put a Dirac operator on any 
(topological) graph while preserving time-reversal symmetry; and the 
assignment of elements in the transition matrix should follow a general 
scheme independent of the particular bonds to which they correspond.  
We saw that difficulties arise due to the fact that, as opposed to the case 
of a Schr\"odinger operator, the Dirac operator is a first order differential 
operator, thus breaking the invariance under $x\mapsto -x$.  In this section 
we show that a quantisation in terms of a Dirac operator can be achieved for 
two component spinors by replacing each bond of the classical (topological) 
graph with a pair of metric bonds one running in each direction.

\begin{figure}[htb]
\begin{center}
\includegraphics[width=4cm]{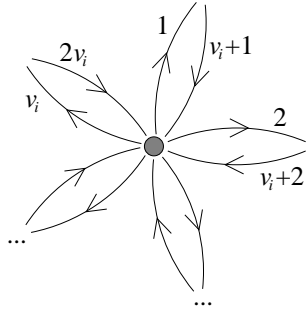}
\end{center}
\caption{An isolated vertex with paired bonds.}
\label{fig:3pair}
\end{figure}

We maintain the realisation of a two component Dirac operator as described 
above, however with the constraint that each vertex $i$ has valency $2v_i$
with pairs of directed bonds meeting at $i$ as indicated in 
figure~\ref{fig:3pair}.  The bonds in a pair replace a single bond of the
given topological graph and hence are assigned the same length
but run in opposite directions.  This restores the physical symmetry of the 
problem as all the original bonds are treated equivalently.  Putting 
plane-wave solutions, equation (\ref{eq:3planewave}), on the bonds 
we can write an eigenfunction on a pair $b$ of bonds as
{\setlength\arraycolsep{2pt}
\begin{equation}\label{eq:3planewavepair}
\begin{split}
\psi_{k}^{b}(x_{b})
&= \mu_{\alpha}^{b} 
   \left( \begin{array}{c} 1 \\ \ui \\ \end{array}\right) 
   \ue^{\ui k x_{b}} + \mu_{\beta}^{b} 
   \left( \begin{array}{c} 1 \\ -\ui \\ \end{array} \right) 
   \ue^{\ui k x_{b}} \\
&\quad + \hat{\mu}_{\alpha}^{b} 
   \left( \begin{array}{c} 1 \\ \ui \\ \end{array}\right) 
   \ue^{-\ui k x_{b}} + \hat{\mu}_{\beta}^{b} 
   \left( \begin{array}{c} 1 \\ -\ui \\ \end{array}\right) 
   \ue^{-\ui k x_{b}} \ ,
\end{split}
\end{equation}
}where $x_{v_{i}+b}$ has been replaced by $L_{b}-x_{b}$ and the extra phases
absorbed into the coefficients.  The coefficients have been labeled 
$\alpha$ or $\beta$ according to the type of spinor, characterised by a lower
component $\ui$ or $-\ui$, rather than the bond on which the spinor travels.
Now let us rearrange the coefficients in the vectors 
$\boldsymbol{\overrightarrow{\mu}}$ and 
$\boldsymbol{\overleftarrow{\mu}}$ so that $\alpha$ coefficients are always 
listed first,
\begin{equation}
\boldsymbol{\overrightarrow{\mu}}=\left( \begin{array}{c}
\mu_{\alpha}^{1}\\
\mu_{\beta}^{1}\\
\vdots \\
\end{array} \right) 
\ , \qquad 
\boldsymbol{\overleftarrow{\mu}}=\left( \begin{array}{c}
\hat{\mu}_{\alpha}^{1}\\
\hat{\mu}_{\beta}^{1}\\
\vdots \\
\end{array} \right) \ .
\end{equation}
The self-adjoint boundary conditions still define a unitary vertex transition 
matrix,
\begin{equation}\label{eq:3verttran2}
\boldsymbol{\overrightarrow{\mu}}=\T^{(i)}\,
\boldsymbol{\overleftarrow{\mu}}\quad \textrm{and}\quad 
(\T^{(i)})^{-1}=(\T^{(i)})^{\dagger} \ .
\end{equation}
Applying the time-reversal operator (\ref{eq:3timeop}) to the 
wavefunctions (\ref{eq:3planewavepair}) we 
find time-reversed vectors 
$\boldsymbol{\overrightarrow{\mu}}_{\mathcal{T}}, \boldsymbol{\overleftarrow{\mu}}_{\mathcal{T}}$,
\begin{equation}
\boldsymbol{\overrightarrow{\mu}}_{\mathcal{T}}=\ui \left( \begin{array}{c}
\overline{\hat{\mu}_{\beta}^{1}}\\
-\overline{\hat{\mu}_{\alpha}^{1}}\\
\vdots \\
\end{array} \right) 
\ , \qquad 
\boldsymbol{\overleftarrow{\mu}}_{\mathcal{T}}=\ui \left( \begin{array}{c}
\overline{\mu_{\beta}^{1}}\\
-\overline{\mu_{\alpha}^{1}}\\
\vdots \\
\end{array} \right) \ .
\end{equation}
Time-reversal invariance again implies
\begin{equation}\label{eq:3timereversec}
\boldsymbol{\overrightarrow{\mu}}_{\mathcal{T}} = \T^{(i)} \, 
\boldsymbol{\overleftarrow{\mu}}_{\mathcal{T}}.
\end{equation}
Using the definition of the transition matrix (\ref{eq:3verttran2}) 
we obtain the following condition for time-reversal invariance,
\begin{equation}\label{eq:3tr2}
(\T^{(i)})^{T}= \left( \begin{array}{ccc}
 J^{-1} & & \\
 & \ddots & \\
 & & J^{-1} \\
\end{array} \right) \T^{(i)} \left( \begin{array}{ccc}
J & & \\
 & \ddots & \\
 & & J \\
\end{array} \right) \ ,
\end{equation}
where
\begin{equation}
J:=
\left( \begin{array}{cc}
0 & 1 \\
-1 & 0 \\
\end{array} \right) \ .
\end{equation}
The form of equation (\ref{eq:3tr2}) is similar to, but not the same as, the 
definition of a symplectic matrix.

The vertex transition matrix $\T^{(i)}$ can be divided into $2\times 2$ 
blocks $(\T^{(i)})^{bc}$ which relate incoming spinors on the pair of 
bonds $b$ to outgoing spinors on the bond pair $c$.  Let 
\begin{equation}
(\T^{(i)})^{bc}
= \left( \begin{array}{cc} 
r & s \\ 
v & w \\ \end{array} \right) \ ,
\end{equation}
then time-reversal symmetry implies 
\begin{equation}  
(\T^{(i)})^{cb} = \left( \begin{array}{cc} 
w & -s \\ 
-v & r \\ 
\end{array} \right) \ ,
\end{equation}
or equivalently,
\begin{equation}
(\T^{(i)})^{cb}= \bigl| (\T^{(i)})^{bc} \bigr| \, 
\left( (\T^{(i)})^{bc} \right) ^{-1} \ .
\end{equation}
This suggests both a method of constructing time-reversal symmetric 
transition matrices and the form of permutation invariance among the bond 
pairs to be required.  Matrices of this form can be written
\begin{equation}\label{eq:3gen}
\T^{(i)}=
\left( \begin{array}{ccc}
u_{1} & & \\
& \ddots & \\
& & u_{v_{i}} \\
\end{array} \right)
\left\{  X \otimes
\left( \begin{array}{cc}
1 & 0 \\
0 & 1 \\
\end{array} \right)
\right\}
\left( \begin{array}{ccc}
 u_{1}^{-1} & & \\
& \ddots & \\
& & u_{v_{i}}^{-1} \\
\end{array} \right) \ ,
\end{equation}
where $X=(x_{bc})$ is a symmetric unitary matrix of dimension $v_{i}$ and 
$u_{j}\in {\mathrm{SU(2)}}$.  $|x_{bc}|^2$ is the probability that any 
incoming state on the classical bond $c$ scatters to any outgoing state 
on the classical bond $b$.  We may now choose to implement a physical 
permutation symmetry between the bonds of the classical graph by requiring 
the matrix $X$ to be invariant under permutations $\sigma$,
\begin{equation}\label{eq:3perm}
X=\sigma X \sigma^{-1} \ .
\end{equation}
Together with the condition that $X$ is unitary this implies
\begin{equation}\label{eq:3X}
X=\ue^{\ui \theta}
\left( \begin{array}{ccc}
q -1 & & q \\
& \ddots & \\
q & & q -1 \\
\end{array} \right) \quad \text{with} \quad 
q :=\frac{1+\ue^{\ui \omega}}{v_{i}} \ .
\end{equation}
Unsurprisingly this is similar to the vertex transition matrix for a 
Schr\"odinger operator 
\cite{paper:kottossmilansky2,paper:kottossmilansky} which is also symmetric
unitary and permutation invariant.  Here the coefficient $\omega$, 
which we can think of parameterising the boundary conditions, is independent 
of $k$.  The extra phase $\ue^{\ui \theta}$ is not interesting as it 
corresponds only to an extra $k$ independent phase factor.  
Adjusting the parameter 
$\omega$ we see that the Dirac equation also allows both Neumann like 
($\omega = 0$) and Dirichlet like ($\omega = \pi$) boundary conditions.  
An element $u_{b}u^{-1}_{c}$ of ${\mathrm{SU(2)}}$ defines a `spinor 
rotation' from the pair of linearly independent two component spinors 
incoming on bond $c$ to the outgoing pair on bond $b$. 

Equations (\ref{eq:3gen}) and (\ref{eq:3X}) define general 
time-reversal invariant vertex transition matrices for two component spinors. 
The transition amplitudes have a form of permutation invariance 
(\ref{eq:3perm}) and in section~\ref{ss:5.1} we 
see that this allows a simple form for the trace formula to be derived.  
With the vertex transition matrix the bond
scattering matrix for the whole graph can be defined as in equation 
(\ref{eq:3S}) with unpaired bonds.  
The quantisation condition (\ref{eq:3quant}) is also unchanged, apart from
the fact that due to doubling of the number of bonds $\SM(k)$ now is
a $4B\times 4B$ matrix.

\section{The Dirac operator for four component spinors}\label{s:4comp}
Before looking at examples for the two component spinors on paired bonds we 
will compare with the alternative construction of the Dirac operator acting 
on four component spinors.  Therefore, we now consider again directed
graphs with $B$ unpaired bonds.  In this case the time independent Dirac 
equation reads
\begin{equation}
-\ui \hbar c \, \alpha \, \frac{\ud \psi}{\ud x} + 
mc^{2}\, \beta \,\psi = E \psi \ ,
\end{equation}
where we choose $\alpha$ and $\beta$ as in (\ref{eq:3dirmat}),
\begin{equation}
\alpha = \left( \begin{array}{cccc}
0 & 0 & 0 & -\ui \\
0 & 0 & \ui & 0 \\
0 & -\ui & 0 & 0 \\
\ui & 0 & 0 & 0 \\
\end{array} \right) 
\ , \qquad
\beta = \left( \begin{array}{cccc}
1 & 0 & 0 & 0 \\
0 & 1 & 0 & 0 \\
0 & 0 & -1 & 0 \\
0 & 0 & 0 & -1 \\
\end{array} \right) \ .
\end{equation}
We could alternatively have chosen either of the other $\alpha$ matrices of 
the standard representation for the Dirac operator in three dimensions or 
used another representation entirely.  The choice here simplifies the working.

Eigenspinors of ${\mathcal D}$ are again plane waves.  For positive energy
they are of the form
\begin{equation}\label{eq:5planewave}
\begin{split}
\psi_{k}^{b}(x_{b})&=\mu_{\alpha}^{b} \left( \begin{array}{c} 
1 \\
0 \\
0 \\
\ui \gamma (k) \\
\end{array} \right) \ue^{\ui k x_{b}} +
 \mu_{\beta}^{b} \left( \begin{array}{c} 
0 \\
1 \\
-\ui  \gamma (k) \\
0 \\
\end{array} \right) \ue^{\ui k x_{b}}  \\ &\quad + 
 \hat{\mu}_{\alpha}^{b} \left( \begin{array}{c} 
1 \\
0 \\
0 \\
-\ui  \gamma (k)\\
\end{array} \right) \ue^{-\ui k x_{b}} +
\hat{\mu}_{\beta}^{b} \left( \begin{array}{c} 
0 \\
1 \\
\ui  \gamma (k)\\
0 \\
\end{array} \right) \ue^{-\ui k x_{b}} \ ,
\end{split}
\end{equation}
with $k>0$. The dispersion relation and $\gamma(k)$ are as in 
(\ref{eq:3dispersion}). These four plane-wave solutions differ from the
two component case although they are labeled with coefficients
$\mu^{b}_{\alpha}$ etc. in analogy with the two component solutions on pairs 
of bonds.  The wavefunction $\psi^{b}(x_{b})$ is a function on a single 
directed bond of a metric graph.  We notice that, in contrast to the pairs of 
two component wavefunctions (\ref{eq:3planewavepair}),
$\psi^{b}(x_{b})$ is not invariant under a change in the bond direction 
$x_{b}\mapsto -x_{b}$.

\subsection{Self-adjoint realisations on graphs}
The construction of self-adjoint realisations of the Dirac operator on 
the graph proceeds in complete analogy to the case of two component
spinors.  We therefore must look for maximal closed subspaces of
\begin{equation}\label{eq:5Sobolevspace}
\bigoplus_{b=1}^B W_{2,1} \bigl( [0,L_b] \bigr) \otimes\kz^4  
\end{equation}
on which the skew-hermitian form 
\begin{equation}\label{eq:4form}
\Omega(\phi,\psi)=\langle \mathcal{D} \phi, \psi \rangle - \langle \phi, 
\mathcal{D} \psi \rangle
\end{equation}
vanishes.  In order to convert the problem to the boundary values of the
spinors we integrate by parts and express $\Omega$ as a complex symplectic 
form,
\begin{equation}\label{eq:4omegascalar}
  \Omega(\phi,\psi)= ( \begin{array}{cc}
  (\boldsymbol{\phi}^{+})^{\dagger}& (\boldsymbol{\phi}^{-})^{\dagger} \\
\end{array} ) \left( \begin{array}{cc}
0 & \UI_{4B} \\
-\UI_{4B} & 0 \\
\end{array} \right) \left( \begin{array}{c}
\boldsymbol{\psi}^{+}\\
\boldsymbol{\psi}^{-}\\
\end{array} \right) \ ,
\end{equation}
on $\kz^{8B}$. Here we have defined
{\setlength\arraycolsep{2pt}
\begin{equation}
\begin{split}
\boldsymbol{\psi}^{+}
&:= \bigl( \psi_{1}^{1}(0),\dots,\psi_{1}^{B}(0),
    \psi_{2}^{1}(0),\dots,\psi_{2}^{B}(0),  \\
&\qquad \psi_{1}^{1}(L_{1}), \dots ,\psi_{1}^{B}(L_{B}),
    \psi_{2}^{1}(L_{1}), \dots ,\psi_{2}^{B}(L_{B}) \bigr)^{T} \ , \\
\boldsymbol{\psi}^{-}
&:= \bigl( -\psi_{4}^{1}(0),\dots,-\psi_{4}^{B}(0), 
   \psi_{3}^{1}(0),\dots,\psi_{3}^{B}(0),  \\
&\qquad \psi_{4}^{1}(L_{1}), \dots ,\psi_{4}^{B}(L_{B}),
   -\psi_{3}^{1}(L_{1}), \dots ,-\psi_{3}^{B}(L_{B}) \bigr)^{T} \ .
\end{split}
\end{equation}
}The map from the spinors on the graph to the vectors $\boldsymbol{\psi}^{+}$ 
and $\boldsymbol{\psi}^{-}$ mixes the first and second components of the 
spinors in $\boldsymbol{\psi}^{+}$ and the third and fourth ones in 
$\boldsymbol{\psi}^{-}$.  Identifying maximally isotropic 
subspaces of $\kz^{8B}$ with respect to $\Omega$, 
equation (\ref{eq:4omegascalar}), the argument proceeds as for the 
two component spinors.  Such a linear subspace is defined by 
\begin{equation}\label{eq:4boundarycond}
\A \boldsymbol{\psi}^{+} + \B \boldsymbol{\psi}^{-} = 0  \ ,
\end{equation}
with $4B\times 4B$ matrices $\A$ and $\B$, and is maximal isotropic if 
$\rk(\A,\B)=4B$ and $\A\B^{\dagger}=\B\A^{\dagger}$.  Under such conditions 
the Dirac operator on (\ref{eq:5Sobolevspace}) with boundary conditions as 
prescribed by (\ref{eq:4boundarycond}) is self-adjoint.

For eigenfunctions of the Dirac operator, i.e. plane-wave solutions
(\ref{eq:5planewave}) on each of the bonds, we define vectors of incoming 
and outgoing coefficients at a single vertex, for example  
figure~\ref{fig:3star},
\begin{equation}
\boldsymbol{\overrightarrow{\mu}}=\left( \begin{array}{c}
\mu^{1}_{\alpha}\\
\mu^{1}_{\beta}\\
\vdots \\
\mu^{w_{i}}_{\alpha}\\
\mu^{w_{i}}_{\beta}\\
\hat{\mu}^{w_{i}+1}_{\alpha}\,\ue^{-\ui kL_{w_{i}+1}}\\
\hat{\mu}^{w_{i}+1}_{\beta}\,\ue^{-\ui kL_{w_{i}+1}}\\
\vdots \\
\hat{\mu}^{v_{i}}_{\alpha}\,\ue^{-\ui kL_{v_{i}}}\\
\hat{\mu}^{v_{i}}_{\beta}\,\ue^{-\ui kL_{v_{i}}}\\
\end{array} \right) \ , \qquad 
\boldsymbol{\overleftarrow{\mu}}=\left( \begin{array}{c}
\hat{\mu}^{1}_{\alpha}\\
\hat{\mu}^{1}_{\beta}\\
\vdots \\
\hat{\mu}^{w_{i}}_{\alpha}\\
\hat{\mu}^{w_{i}}_{\beta}\\
\mu^{w_{i}+1}_{\alpha}\,\ue^{\ui kL_{w_{i}+1}}\\
\mu^{w_{i}+1}_{\beta}\,\ue^{\ui kL_{w_{i}+1}}\\
\vdots \\
\mu^{v_{i}}_{\alpha}\,\ue^{\ui kL_{v_{i}}}\\
\mu^{v_{i}}_{\beta}\,\ue^{\ui kL_{v_{i}}}\\
\end{array} \right) \ .
\end{equation}
Then using boundary conditions defined at the vertex by matrices $\A^{(i)}$ 
and $\B^{(i)}$ we find as in the two component case,
\begin{equation}
\boldsymbol{\overrightarrow{\mu}}=-(\A^{(i)} - \ui \gamma (k) \B^{(i)})^{-1}
(\A^{(i)} + \ui \gamma (k) \B^{(i)})\, 
\boldsymbol{\overleftarrow{\mu}} \ .
\end{equation}
The vertex transition matrix 
\begin{equation}\label{eq:4transition}
 \T^{(i)}:=-(\A^{(i)} - \ui \gamma (k) \B^{(i)})^{-1}
 (\A^{(i)} + \ui \gamma (k) \B^{(i)})
\end{equation}
is unitary due to the condition 
$\A^{(i)}\B^{(i)\dagger}=\B^{(i)}\A^{(i)\dagger}$ that guarantees current 
conservation at the vertex.

\subsection{Time-reversal symmetry}
We now turn to see what time-reversal invariance implies for the vertex 
transition matrix.  
The time-reversal operator in the standard representation is 
\begin{equation}
\mathcal{T}=-
\left( \begin{array}{cccc}
0 & -\ui & 0 & 0 \\
\ui & 0 & 0 & 0 \\
0 & 0 & 0 & -\ui \\
0 & 0 & \ui & 0 \\
\end{array} \right) \mathcal{K} \ .
\end{equation}
This commutes with the Hamiltonian on the bonds so there is no requirement 
that the mass be zero with four component spinors.  Applying $\mathcal{T}$ 
to the wavefunction at the vertex $i$ we define new vectors from the 
coefficients of outgoing and incoming waves, 
$\boldsymbol{\overrightarrow{\mu}}_{\mathcal{T}}, 
\boldsymbol{\overleftarrow{\mu}}_{\mathcal{T}}$,
\begin{equation}
\boldsymbol{\overrightarrow{\mu}}_{\mathcal{T}} = \ui 
\left( \begin{array}{ccc} 
J
& & \\
& \ddots & \\
 &  & J\\
\end{array} \right) \overline{(\boldsymbol{\overleftarrow{\mu}})} \ ,\quad
%
\boldsymbol{\overleftarrow{\mu}}_{\mathcal{T}} = \ui 
\left( \begin{array}{ccc} 
J
& & \\
& \ddots & \\
 &  & J\\
\end{array} \right) \overline{(\boldsymbol{\overrightarrow{\mu}})} \ .
\end{equation}
%
%
For time-reversal invariance we require
\begin{equation}\label{eq:4tri}
\boldsymbol{\overrightarrow{\mu}}_{\mathcal{T}}= \T^{(i)} \,  
\boldsymbol{\overleftarrow{\mu}}_{\mathcal{T}} \ .
\end{equation}
Using the unitarity of $\T^{(i)}$ time-reversal symmetry 
(\ref{eq:4tri}) is equivalent 
to demanding the transition matrix satisfy
\begin{equation}\label{eq:4tr2}
(\T^{(i)})^{T}= \left( \begin{array}{ccc}
 J^{-1} & & \\
 & \ddots & \\
 & & J^{-1} \\
\end{array} \right) \T^{(i)} \left( \begin{array}{ccc}
J & & \\
 & \ddots & \\
 & & J \\
\end{array} \right) \ .
\end{equation}
This is the same as condition (\ref{eq:3tr2}) derived for two component 
spinors on paired bonds.  The general form of transition matrices 
(\ref{eq:3gen}) proposed for the spinors on paired bonds will hence also 
hold for four component spinors on a directed graph.  Furthermore with  
four component spinors the mass is no longer required to be zero.  Instead 
any $\T^{(i)}$ which can be constructed from the boundary conditions 
according to equation (\ref{eq:4transition}) and which is also time-reversal 
invariant is available.

\section{Energy level statistics}\label{s:stats}
To verify the conjecture of Bohigas, Giannoni and Schmit in the case of the
time-reversal invariant Dirac operator on a graph we examine 
the energy level statistics through both numerical calculations and directly
via the trace formula.

From now on we consider only boundary conditions with the form 
of general vertex transition matrices defined in equations 
(\ref{eq:3gen}) and (\ref{eq:3X}).  This still leaves a range of boundary 
conditions parameterised by $\omega$.  Of particular interest are the 
Neumann like boundary conditions, $\omega=0$.  These are the boundary 
conditions most often studied for the Schr\"odinger operator on the graph.
The vertex transition matrix $\T^{(i)}$ 
for Neumann boundary conditions is generated by 
the matrices
\begin{equation}
\A^{(i)}=U^{(i)}\left\{ \left( \begin{array}{ccccc}
1 & -1 & 0 & 0 & \dots \\
0 & 1 & -1 & 0 & \dots \\
& & \ddots & \ddots & \\
0 & \dots & 0 & 1 & -1 \\
0  & \dots & 0& 0 & 0 \\
\end{array} \right) \otimes \left( \begin{array}{cc}
1 & 0 \\
0 & 1 \\ 
\end{array} \right) \right\} (U^{(i)})^{-1} \ ,
\end{equation}
\begin{equation}
\B^{(i)}= U^{(i)}\left\{ \left( \begin{array}{cccc}
0 & 0 &\dots & 0 \\
\vdots & \vdots & & \vdots \\
0 & 0 &\dots & 0 \\
1 & 1 &\dots & 1 \\
\end{array} \right) \otimes \left( \begin{array}{cc}
1 & 0 \\
0 & 1 \\ 
\end{array} \right) \right\} (U^{(i)})^{-1} \ ,
\end{equation}
where $U^{(i)}$ is a block diagonal matrix of elements of ${\mathrm{SU(2)}}$,
\begin{equation}\label{eq:4Umatrix}
U^{(i)}=\left( \begin{array}{ccc}
u_{1} & & \\
& \ddots & \\
& & u_{v_{i}} \\
\end{array} \right) \ .
\end{equation}
Substituting these boundary conditions into equation (\ref{eq:4transition}) 
and using some algebra we obtain the vertex transition matrix
\begin{equation}\label{eq:4Tneumann}
\T^{(i)}=U^{(i)}\left\{ \left( \begin{array}{ccc}
\frac{2}{v_{i}}-1 & & \frac{2}{v_{i}} \\
 &\ddots &  \\
\frac{2}{v_{i}} &  & \frac{2}{v_{i}}-1 \\
\end{array} \right) \otimes \left( \begin{array}{cc}
1 & 0 \\
0 & 1 \\ 
\end{array} \right) \right\} (U^{(i)})^{-1} \ .
\end{equation}
$\T^{(i)}$ is independent of the function $\gamma(k)$ determined by the mass.
With this form of the Neumann boundary conditions there is no need to specify 
the mass or whether the system is represented by paired two component spinors 
or four component spinors on a directed graph.

\begin{figure}[htb]
\begin{center}
\includegraphics[width=14cm]{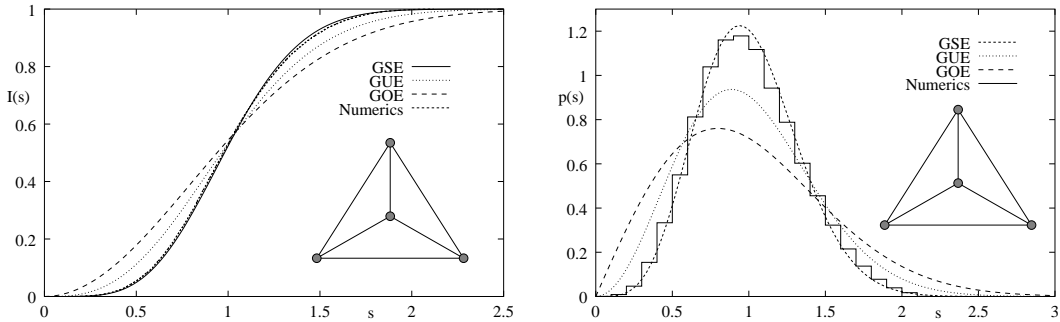}
\end{center}
\caption{The integrated nearest neighbour spacing distribution and 
a histogram of the spacing distribution for a Dirac operator with Neumann 
boundary conditions on a fully 
connected square, 24234 levels.}
\label{fig:5fcs}
\end{figure}

Figure \ref{fig:5fcs} shows the nearest neighbour spacing statistics for a 
fully connected square with Neumann boundary conditions, the vertex 
transition matrix defined in equation (\ref{eq:4Tneumann}).  
The three elements $u_{b}\in {\mathrm{SU(2)}}$ at a vertex 
were chosen as
\begin{equation}
u_{b}=\exp( \ui \theta_{b} \sigma_{n_b}) \ ,
\end{equation}
where $\sigma_{n_b}$ is a Pauli matrix.  The parameters $\theta_{b}$ 
were then selected randomly at each vertex.  
To obtain good agreement with the symplectic random matrix 
ensemble the parameters were also picked such that the elements 
$u_{b}$ generated at the vertices were all sufficiently different.  
Deviations from the GSE behaviour become apparent if the elements of 
${\mathrm{SU(2)}}$ at one vertex are similar 
to those at another, i.e. $\theta_{b}^{l} \approx \theta_{c}^{m}$.

\begin{figure}[htb]
\begin{center}
\includegraphics[width=14cm]{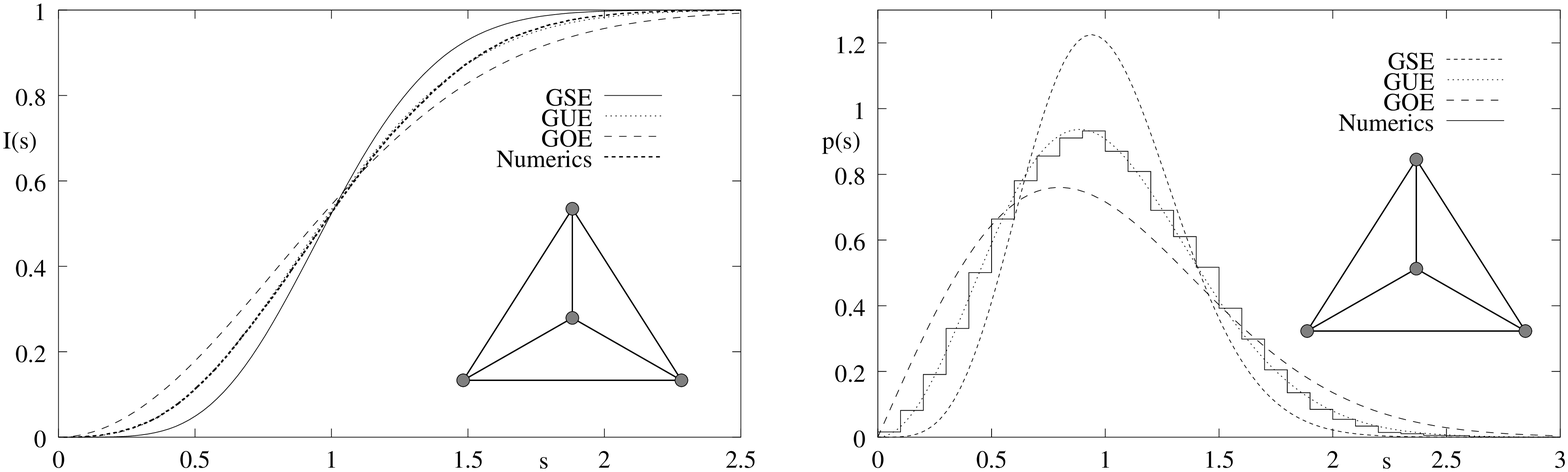}
\end{center}
\caption{The integrated nearest neighbour spacing distribution and 
a histogram of the spacing distribution for a Dirac operator with Neumann 
boundary conditions on a fully 
connected square where time reversal symmetry has
been broken with a magnetic vector potential, 25503 levels.}
\label{fig:5ntrsfcs}
\end{figure}

Introducing a magnetic vector potential $A_{b}$ on the bonds breaks 
time-reversal symmetry.  The transition elements for the $S$-matrix remain 
the same but a plane wave propagating down a bond $(i\rightarrow j)$ picks
up an extra phase $\exp(\ui A_{(ij)} L_{(ij)})$ where we know that 
$L_{(ji)}=L_{(ij)}$ but $A_{(ji)}=-A_{(ij)}$.  Figure \ref{fig:5ntrsfcs} 
shows that for the Dirac operator on a fully connected square breaking the
time-reversal invariance in this way produces energy level statistics 
like those of random GUE matrices as expected.  For the Dirac operator we 
used Neumann boundary conditions and random elements of ${\mathrm{SU(2)}}$ 
chosen as previously.

\subsection{Trace formula}\label{ss:5.1}
A trace formula for the Laplacian on a graph was first produced by Roth 
\cite{paper:roth}.
We derive the trace formula for the density of states of a system with 
Neumann boundary conditions, following the approach developed by Kottos 
and Smilansky \cite{paper:kottossmilansky2,paper:kottossmilansky}
for the Schr\"odinger operator.  The vertex transition matrices 
for such a system were defined in equation (\ref{eq:4Tneumann}).  The main
tool in setting up the trace formula is the bond scattering matrix
$\SM(k)$ defined in equation (\ref{eq:3S}).

Consider the density of states for the wave-number $k$.  It is 
an infinite series of delta functions located at $k=k_n$ corresponding
to eigenvalues $E_n$ of the Dirac operator.  The eigenvalues are counted
according to their multiplicity.
Wave-numbers $k_n$ are values of $k$ for which the function 
\begin{equation}
\zeta(k):=|\UI_{4B}-\SM(k)|
\end{equation}
is zero.  Diagonalising $\SM(k)$ we can write $\zeta(k)$ in terms of the 
eigenvalues $\ue^{\ui \phi_{j}(k)}$ of $\SM(k)$,
%
\begin{equation}
|\UI_{4B}-\SM(k)|=\prod_{j=1}^{4B}(1-\ue^{\ui \phi_{j}(k)})
= 2^{4B} |\SM(k)|^{\frac{1}{2}} \prod_{j=1}^{4B} 
\sin \left( \frac{\phi_{j}(k)}{2} \right) \  . 
\end{equation}
%
$\zeta(k)$ is complex with the phase contained in the term 
$|\SM(k)|^{\frac{1}{2}}$.  Multiplying by $|\SM(k)|^{-\frac{1}{2}}$ we 
define a real function whose zeros are also the zeros of $\zeta(k)$.  
Then the density of states can be written
\begin{equation}\label{eq:5dos}
d(k) = \sum_{n=1}^\infty \delta(k-k_n) = 
-\frac{1}{\pi}  \lim_{\epsilon \rightarrow 0}\, \textrm{Im}\frac{\ud}{\ud k} 
\log (|\UI_{4B}-\SM(k+\ui \epsilon)| |\SM(k+\ui \epsilon)|^{-\frac{1}{2}}) \ .
\end{equation}
The expression (\ref{eq:5dos}) factorises into a sum of two terms.  
The smooth part of the spectral density is
{\setlength\arraycolsep{2pt}
\begin{eqnarray}\label{eq:5dsmooth}
d_{\textrm{smth}}(k) & := &
-\frac{1}{\pi}  \lim_{\epsilon \rightarrow 0}\, \textrm{Im}
\frac{\ud}{\ud k} \log
(|\SM(k+\ui \epsilon)|^{-\frac{1}{2}}) \nonumber \\
& = & \frac{1}{2\pi} \textrm{Im} \frac{\ud}{\ud k} \log
(\ue^{\ui 4kL}) \nonumber \\
& = & \frac{2L}{\pi} \ , 
\end{eqnarray}
}where $L$ is the total length of the graph,
\begin{displaymath}
 L=\sum_{b=1}^{B} L_{b} \ .
\end{displaymath}
(Note that in our definition of $L$ the length of each bond of the classical
graph is counted only once although the S-matrix can be considered either 
as scattering between two component spinors on paired bonds or four component 
spinors on a directed graph.)

The oscillating part of the spectral density is
\begin{equation}\label{eq:5dosc}
d_{\textrm{osc}}(k):=
-\frac{1}{\pi} \lim_{\epsilon \rightarrow 0} \textrm{Im}\frac{\ud}{\ud k}
\log (|\UI_{4B}-\SM(k+\ui\epsilon)|)\ .
\end{equation}
The logarithm can be expanded as a sum of powers of the trace of $\SM(k)$,
\begin{equation}\label{eq:5trace}
\log (|\UI_{4B}-\SM(k)|)=-\sum_{n=1}^{\infty} \frac{1}{n} 
\tr \SM^{n}(k) \ .
\end{equation}
As in the case of the Schr\"odinger operator on a graph we can now write 
the trace of powers of $\SM(k)$ as a sum over periodic orbits on the graph.  
However, for the Dirac operator, the transition amplitudes between 
classical bonds are specified by the $2\times 2$ matrices $\T^{bc}$,
\begin{equation}
\tr \SM^{n}(k)= \tr \sum_{b_{1}\dots b_{n}} \ue^{\ui k L_{b_{1}}}
 \T^{b_{1}b_{2}} \ue^{\ui k L_{b_{2}}} 
 \T^{b_{2}b_{3}} \dots 
\ue^{\ui k L_{b_{n}}} \T^{b_{n}b_{1}} \ .
\end{equation}
Thus $\tr \SM^{n}(k) \ne 0$ implies that the sequence 
$(b_{1}b_{2}\dots b_{n})$ of bonds labels a periodic orbit on the graph.
For a single periodic orbit $p$ which passes through $n$ vertices,
\begin{equation}\label{eq:5spo}
 \tr ( \T^{b_{1}b_{2}} \dots \T^{b_{n}b_{1}})=
A_{p}\, \textrm{tr}( d_{p})\, \ue^{\ui\pi\mu_p} \ .
\end{equation}
In this formula $A_{p}$ is a (positive) stability factor, the product of 
the absolute values of the transition amplitudes for spinor pairs round 
the orbit,
\begin{equation}
A_{p}:=\prod_{s=1}^{\mu_{p}}\left( 1-\frac{2}{v_{s}} \right) 
\prod_{t=1}^{\nu_{p}}
\left( \frac{2}{v_{t}} \right) \ ,
\end{equation}
where $\mu_{p}$ is the number of cases of back scattering from vertices with 
$v_{s} > 2$ and $n=\mu_{p}+\nu_{p}$.  An overall sign is collected in the 
factor $\ue^{\ui\pi\mu_p}$ and $d_p$ is the product of the elements of 
${\mathrm{SU(2)}}$ that rotate between the two types of spinors on the 
periodic orbit,
\begin{equation}
d_p := u^{b_{1}b_{2}}u^{b_{2}b_{3}}\dots u^{b_{n}b_{1}} \ .
\end{equation}
The spin rotations $u^{b_{i}b_{i+1}}:=u_{b_{i+1}}u_{b_{i}}^{-1}$ 
are composed of ${\mathrm{SU(2)}}$-blocks comprising the matrix $U^{(i)}$,
see equation (\ref{eq:4Umatrix}), attached to the $i$th vertex visited 
along the periodic orbit. 

Combining these results,
\begin{equation}\label{eq:5S^n}
\tr \SM^{n}(k) = \sum_{p \in P_{n}}  
A_{p}\, \ue^{\ui\pi\mu_p}\, \tr( d_p )\, \ue^{\ui k l_{p}} \ ,
\end{equation}
where the sum is over the set $P_{n}$ of periodic orbits of $n$ bonds and 
$l_{p}$ is the metric length of the orbit $p$.  Substituting equations 
(\ref{eq:5S^n}) and (\ref{eq:5trace}) into equation (\ref{eq:5dosc}) we find 
the oscillating part of the density of states.  Adding the result for the 
smooth part (\ref{eq:5dsmooth}) we obtain
\begin{equation}\label{eq:5d}
d(k)=\frac{2L}{\pi}+ \frac{1}{\pi} \sum_{p} \frac{l_{p}}{r_{p}}   
A_{p}\, \ue^{\ui\pi\mu_p} \, \tr(d_{p}) \cos (k l_{p}) \ .
\end{equation}
The sum is over all periodic orbits.  If an orbit $p$ consists of 
$r_{p}$ repetitions 
of a shorter orbit then only the length of the primitive orbit 
$l_{p}/r_{p}$ is included in the formula. This trace formula provides 
an exact relation for the density of states of the Dirac operator 
on a graph with Neumann boundary conditions as a sum over the classical 
periodic orbits.

Equation (\ref{eq:5d}) for the density of states can be compared both to 
the form derived for the Schr\"odinger operator on a graph
\cite{paper:roth,paper:kottossmilansky2,paper:kottossmilansky} and to previous 
semiclassical results for systems with spin 
\cite{paper:boltekeppler, paper:boltekeppler2, paper:kepplermarklofmezzadri}.
Most terms in our density of states are the same as those derived by Kottos 
and Smilansky for the Schr\"odinger operator 
\cite{paper:kottossmilansky2,paper:kottossmilansky}.  
The additional term for the Dirac operator is the trace of $d_{p}$.  This 
can be thought of as an additional weight factor which depends on the 
transformation of the spinor pairs around the orbit.  

Comparing our trace formula to the general semiclassical form for the Dirac 
operator \cite{paper:boltekeppler, paper:boltekeppler2},
we see that here the Dirac operator also produces a trace over an element of
${\mathrm{SU(2)}}$ associated with the change in spin around the orbit.  
This weight factor can also be interpreted in terms of an effective 
rotation angle of a classical spin vector transported along the periodic
orbit according to the equation of Thomas precession.
That the results agree is not unexpected.  The Dirac operator on a graph 
however does not require a semiclassical approximation to derive
the density of states.  Finally we also note the similarity with the 
trace formula for a different system, the quantum cat map with spin, derived 
by Keppeler, Marklof and Mezzadri, \cite{paper:kepplermarklofmezzadri}.  
In this case the quantum map is an equivalent to the Pauli operator with
spin 1/2.  
In their semiclassical trace formula the presence of spin also appears as 
a trace of an ${\mathrm{SU(2)}}$ matrix which defines the spin transport 
around a periodic orbit.

\subsection{Form factor}\label{ss:5.2}
From the trace formula we derive the spectral two-point form factor for the 
Dirac operator on quantum graphs.  The form factor is an energy level 
statistic that is widely studied for classically chaotic quantum systems and 
in particular for the Schr\"odinger operator on graphs, see 
\cite{paper:kottossmilansky2,paper:kottossmilansky,paper:berkolaikokeating,paper:berkolaikoschanzwhitney,paper:berkolaikoschanzwhitney2}.  
We see that our form factor agrees with the general semiclassical result 
for systems with spin \cite{paper:boltekeppler3}, and   
making the diagonal approximation the form factor 
for low $\tau$ is consistent with the GSE prediction from random matrix theory.

We derive the form factor for the energy level spacings.  
This is the 
Fourier transform of the two-point correlation function itself defined as
\begin{equation}
R_{2}(x) := \left( \frac{\pi}{2L} \right)^{2} \lim_{\Lambda \rightarrow 
\infty} \frac{1}{\Lambda} \int_{0}^{\Lambda} d(k)\,d\left( k- \frac{\pi x}{L}
\right) \ud k -1 \ .
\end{equation}
In defining the two-point correlation function we have rescaled the spectrum 
dividing by the mean level spacing $2L/\pi$, see equation 
(\ref{eq:5dsmooth}).  
For systems with half-integer 
spin and time-reversal invariance the eigenvalues $k_{n}$ come in pairs 
(Kramers' degeneracy).  We consider the spectrum $x_{n}$ where this degeneracy 
has been lifted, each pair of eigenvalues being replaced with a single 
representative.  The mean spacing of the $x_{n}$ distribution is therefore 
$L/\pi$. Substituting the trace formula for the density of states  
(\ref{eq:5d}) into the definition of $R_{2}(x)$ and carrying out the integral 
we find 
\begin{equation}\label{eq:5r2}
R_{2}(x) = \frac{1}{2(2L)^2}  
\sum_{p,q} \frac{l_{p}l_{q}}{r_{p}r_{q}} A_{p} A_{q} \, 
\ue^{\ui\pi(\mu_p +\mu_q)}
\tr (d_{p}) \tr (d_{q}) \,
\cos \left( \frac{\pi x l_{q}}{L} \right) \delta_{l_{p},l_{q}} \ .
\end{equation}

The two-point form factor $K_{2}(\tau)$ is the Fourier transform of the 
two-point correlation function,
\begin{equation}
K_{2}(\tau) := 
 \int_{- \infty}^{\infty} R_{2}(x)\,\ue^{2\pi \ui x \tau} \ud x \ ,
\end{equation}
which is even in $\tau$.
Taking the Fourier transform of $R_{2}(x)$ term-wise we obtain
\begin{equation}\label{eq:5K2}
K_{2}(\tau) = \frac{1}{4(2L)^2}  
\sum_{p,q} \frac{l_{p}l_{q}}{r_{p}r_{q}} A_{p} A_{q} \, 
\ue^{\ui\pi(\mu_p +\mu_q)} \tr (d_{p}) \tr (d_{q}) \,
\delta \left( \tau  - \frac{l_{q}}{2L} \right) \delta_{l_{p},l_{q}}
\end{equation}
for $\tau$ positive.

The eigenvalue distribution of random matrices averaged over the GSE 
has a two-point form factor 
\begin{equation}\label{eq:5K2gse}
K_{2}^{GSE}(\tau) = \left\{ \begin{array}{lcl}
\frac{1}{2} |\tau | -\frac{1}{4} |\tau | \log |1- |\tau| | & &
\textrm{for } |\tau | \le 2 \\
1 & & \textrm{for } |\tau | \ge 2 \\
\end{array} \right. \ .
\end{equation}
Making similar assumptions to those often used in the Schr\"odinger case 
we will see that $K_{2}(\tau)\sim |\tau|/2$ for $|\tau|\to 0$. 
This agrees with the random matrix theory prediction of equation 
(\ref{eq:5K2gse}).

The analysis of the form factor based on the trace formula requires us to
take a semiclassical limit.  The latter can be characterised by going to
some regime with increasing mean spectral density.  According to the
relation (\ref{eq:5dsmooth}) this corresponds to increasing the total
length $L$ of the bonds.  We realise this limit in passing to graphs
with an increasing number $B$ of bonds while keeping the mean bond length
$\overline{L}:=L/B$ constant.    
As $\tau\to 0$ the form factor should be dominated by the
diagonal form factor introduced by Berry \cite{paper:berry},
\begin{equation}\label{eq:5K2-1}
K_{2}^{diag}(\tau) := \frac{1}{4(2L)^2}  
\sum_{p} \left( \frac{l_{p}}{r_{p}} \right)^{2} 
\left\{ A_{p}^{2} (\tr (d_{p}))^{2} + 
A_{p} A_{\overline{p}} 
\tr (d_{p}) \tr (d_{\overline{p}}) \right\}
\delta \left( \tau  - \frac{l_{p}}{2L} \right) \ ,
\end{equation}
since in this limit only correlations for an orbit $p$ with itself and 
with its time-reversed partner $\overline{p}$ are relevant. Combined with
the semiclassical limit we now have to consider $L\to\infty$, $\tau\to 0$
with $L\tau\to \infty$.

Comparing $p$ and $\overline{p}$ we see that $A_{p}=A_{\overline{p}}$ and 
$d_{\overline{p}}=d_{p}^{-1}$.  For elements of ${\mathrm{SU(2)}}$  
$\tr (d^{-1})= \tr (d)$,
\begin{equation}\label{eq:5K2-2}
K_{2}^{diag}(\tau) = \frac{1}{2(2L)^2}  
\sum_{p} \left( \frac{l_{p}}{r_{p}} \right)^{2} 
A_{p}^{2}\, (\tr (d_{p}))^{2} \,
\delta \left( \tau  - \frac{l_{p}}{2L} \right) \ .
\end{equation}
In the limit $L\tau \rightarrow \infty $ it is the long orbits which dominate 
the sum and for these the proportion which are repetitions of shorter 
orbits tends to zero so we ignore periodic orbits with $r_{p} \ne 1$.
Breaking the sum on periodic orbits into a sum on sets $P_{n}$ of orbits 
of $n$ bonds and using $l_{p} \approx n \overline{L}$,
\begin{equation}\label{eq:5K2-3}
K_{2}^{diag}(\tau) \approx \frac{1}{2}  
\sum_{n} \left( \frac{n}{2B} \right)^{2} \delta 
\left( \tau  - \frac{n}{2B} \right) \sum_{p \in P_{n}} 
A_{p}^{2}\, (\tr (d_{p}))^{2} \ . 
\end{equation}

We can regard both $A_{p}$ and $d_{p}$ as functions of $p$.  The elements of 
${\mathrm{SU(2)}}$ in $d_{p}$ were selected randomly, according to Haar
measure, so $A_{p}$ and $d_{p}$ are uncorrelated.  If the number of periodic 
orbits of length $n$ is $|P_{n}|$ then for large $n$
\begin{equation}\label{eq:5uncorrelated}
  \frac{1}{|P_{n}|} \sum_{p \in P_{n}} A_{p}^{2} (\tr (d_{p}))^{2} \sim
  \left( \frac{1}{|P_{n}|} \sum_{p \in P_{n}} A_{p}^{2} \right)
  \left( \frac{1}{|P_{n}|} \sum_{p \in P_{n}} (\tr (d_{p}))^{2} \right) \ .
\end{equation}
We now consider the two averages separately.  The sum over $P_{n}$ of 
$A_{p}^{2}$ can be evaluated in the limit $n\rightarrow \infty$ using the 
classical ergodicity of the graph.  Classical motion on the graph is 
determined by the doubly stochastic matrix $M=(M_{bc})$, with the magnitude 
squared of the transition amplitudes between the bonds as entries,
\begin{equation}
M_{bc}:=|X_{bc}|^{2} \ .
\end{equation}
$M$ defines a Markov chain.  For a connected graph a given pair $(b,c)$ 
of bonds can be linked by a path of $q$ bonds.  Thus $(M^q)_{bc}>0$ such 
that $M$ is an irreducible matrix and the Markov chain is ergodic.  
If, in addition, no eigenvalue other than one lies on the unit circle 
it is mixing which implies 
\begin{equation}
\lim_{n\rightarrow \infty}  (M^{n})_{bc} = \frac{1}{2B} \ . 
\end{equation}
The trace of $M^{n}$ therefore approaches unity, but the trace can also be 
expressed as a sum over the periodic orbits,
\begin{equation}
\tr M^{n} = n \sum_{p \in P_{n}} A_{p}^{2} \ .
\end{equation}
The factor $n$ counts the cyclic permutations of the orbit.  Thus in the 
limit of long orbits we may replace $\sum_{p \in P_{n}} A_{p}^{2}$ with 
$1/n$ if the Markov process is mixing.  It should be noted that while we 
have used the mixing property of the classical motion on the graph without 
spin (defined by the matrices $X$) it is equivalent to the same property 
of a $4B$ dimensional matrix $M$ defined from $\T$.  In this case 
$M_{bc}=|T_{bc}|^{2}$ and each spinor is treated separately.  We defined 
the mixing property of the classical graph using $X$ to make the connection 
with the amplitudes $A_{p}$ clear.

In the limit of $n \rightarrow \infty$ we replace the average over $P_{n}$ of 
$(\tr (d_{p}))^{2}$ with the integral over ${\mathrm{SU(2)}}$ with Haar 
measure.  Evaluating the integral directly we obtain
\begin{equation}
\lim_{n\to\infty}\frac{1}{|P_{n}|} \sum_{p \in P_{n}} (\tr (d_{p}))^{2}
= \int_{\mathrm{SU(2)}} (\tr (u))^{2} \ud u = 1 \ .
\end{equation}
Substituting these results into the diagonal approximation for the form factor 
(\ref{eq:5K2-3}) we find 
\begin{equation}\label{eq:5K2-4}
K_{2}^{diag}(\tau) \sim \frac{1}{2}  
 \sum_{n} \frac{1}{2B} \,\frac{n}{2B} \delta 
\left( \tau  - \frac{n}{2B} \right)\sim \frac{1}{2} |\tau |  
\end{equation}
in the combined limit $L\to\infty$, $\tau\to 0$ with $L\tau\to \infty$.
This is in agreement with the GSE prediction for small $|\tau|$.  
The validity of the 
assumptions made in the argument is of less consequence than that they are of
the same type used in other semiclassical approximations, particularly for the
Schr\"odinger operator on a graph.  The key additional assumption made to 
include the spin evolution was equation (\ref{eq:5uncorrelated}),  
taking the elements $d_{p}$ of ${\mathrm{SU(2)}}$ to be uncorrelated 
with the weights $A_{p}$ of the orbit.

\begin{figure}[htb]
\begin{center}
\includegraphics[width=8cm]{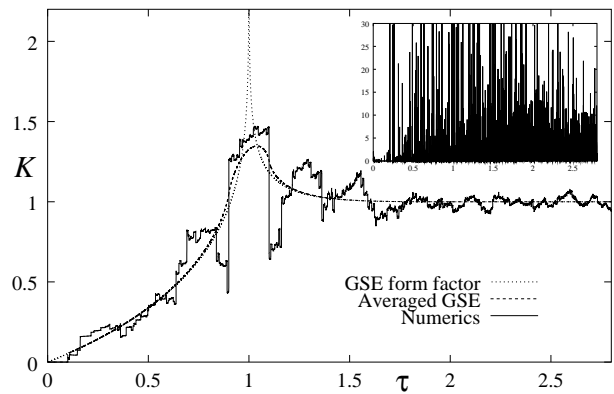}
\end{center}
\caption{Plots of an averaged spectral form factor from the fully connected
square and the GSE prediction.  The inset shows the result for the form 
factor without averaging.}
\label{fig:formfactor}
\end{figure}

Figure \ref{fig:formfactor} compares a numerical calculation of the form 
factor with the GSE prediction (\ref{eq:5K2gse}).  As the form 
factor takes the form of a distribution, see equation (\ref{eq:5K2}), 
meaningful
results are only obtained when it is averaged with a test function.  For 
simplicity we use a function which is one in the region 
$\tau \pm 0.1$ and zero outside.  The theoretical prediction for the GSE form 
factor can also be averaged in the same way.  As the computation of the form 
factor leads to a delta function at $\tau=0$ in the limit of the number of 
levels going to infinity the first few values of $\tau$
 are omitted in the numerical calculation.

\section{Gedanken experiment}\label{s:exp}
It is common when studying graphs, in this context, to describe them as an 
idealised model of a ``typical'' quantum system.  When compared to 
quantum maps or systems of constant negative curvature the case is certainly 
strong.  However in introducing the Dirac operator to graphs we may have 
inadvertently weakened the argument and that is what we seek to correct here.

\begin{figure}[htb]
\begin{center}
\includegraphics[width=3cm]{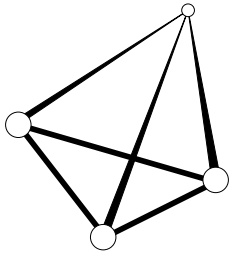}
\end{center}
\caption{A graph realised as a network of wires embedded in $3d$.}
\label{fig:6gedanken}
\end{figure}

Let us think of the classical graph that we wish to quantise as a network or
web of wires set up in our ideal laboratory, see figure \ref{fig:6gedanken}.
To quantise this system using the Dirac operator we must introduce 
pairs of spinors propagating in both directions along the wires.  At the 
vertices changes in the composition of the spinor pairs are described 
by elements $u_{bc} \in {\mathrm{SU(2)}}$, $c$ labels the incoming classical 
bond and $b$ the outgoing bond.  These elements $u_{bc}$ necessarily 
depend on the pair of wires under consideration, $u_{bc} \ne u_{de}$.
This is not a feature
of the usual Schr\"odinger quantisation 
\cite{paper:kottossmilansky2,paper:kottossmilansky}
where all wires meeting at a vertex are treated equivalently.  It is 
therefore fair to
ask whether our model web could really distinguish bonds in such a way.

One simple geometric argument connecting the matrices $u_{bc}$ to our 
experimental network is provided by the double covering map 
$f:{\mathrm{SU(2)}} \rightarrow {\mathrm{SO(3)}}$.
Let $R(\theta_{bc})\in {\mathrm{SO(3)}}$ describe the rotation from bond 
$c$ to $b$ at a vertex.  Setting $u_{bc}=f^{-1}(R(\theta_{bc}))$, with
some choice of the branch $f^{-1}({\mathrm{SO(3)}})$, we define 
a set of elements of ${\mathrm{SU(2)}}$ from the architecture of the network.  
This construction is time-reversal symmetric as $u_{cb}=u_{bc}^{-1}$.  
(To write this in the form used for the vertex boundary conditions 
(\ref{eq:4Tneumann}) choose a reference direction at each 
vertex and take the $v$ 
elements $u_{c}$ to be $f^{-1}(R(\theta_{c}))$ where $R(\theta_{c})$ rotates 
from the reference direction to bond $c$.)
We have not attempted to provide a physical justification for this 
transformation of spinor pairs at the vertices.  Our formulation rather shows
that the extent to which bonds must be distinguished when quantising a graph 
with the Dirac operator remains consistent with the picture of the graph
as a simple ideal quantum system.  There is sufficient geometrical information 
in the structure of a graph embedded in three dimensions to determine 
rotations of the spinors at the vertices.

\subsection*{Acknowledgments}
We would like to thank Stefan Keppeler for useful discussions and 
Arnd B\"acker for advice with the numerics. 
This work has been fully supported by the European Commission under the 
Research Training Network (Mathematical Aspects of Quantum Chaos) 
no. HPRN-CT-2000-00103 of the IHP Programme.

\bibliography{../reffs/papers.bib,../reffs/books.bib}

\end{document}